\documentstyle[preprint,prd,aps]{revtex}
\tightenlines
\input epsf
\def\kp{k_\perp}
\def\ee{e^+e^-}
\def\HW{{\small HERWIG}}
\def\VEV#1{\langle #1\rangle}
\def\re#1{(\ref{#1})}
\def\beq{\begin{equation}}   \def\eeq{\end{equation}}
\def\beeq{\begin{eqnarray}}   \def\eeeq{\end{eqnarray}}
\begin{document}
\preprint{Cavendish--HEP--98/10}
\title{Comparisons of New Jet Clustering Algorithms for
Hadron-Hadron Collisions}
\author{A.T.\ Pierce\thanks{Present address: Department of Physics,
University of California, Berkeley, CA 94720-7300, USA.}
and B.R.\ Webber\thanks{Research
supported in part by the U.K. Particle Physics and Astronomy Research
Council and by the E.U.\ Fourth Framework Programme `Training and Mobility of
Researchers', Network `Quantum Chromodynamics and the Deep Structure of
Elementary Particles', contract FMRX-CT98-0194 (DG 12 - MIHT).}}
\address{Cavendish Laboratory, University of Cambridge,\\
        Madingley Road, Cambridge CB3 0HE, U.K.\\
        E-mail: {\tt apierce@fnal.gov}, {\tt brw1@cam.ac.uk}}
\date{\today}
\maketitle
\begin{abstract}
We investigate a modification of the $\kp$-clustering jet
algorithm for hadron-hadron collisions, analogous to the ``Cambridge''
algorithm recently proposed for $\ee$ annihilation, in which pairs of
objects that are close in angle are combined preferentially. This has the
effect of building jets from the inside out, with the aim of
reducing the amount of the underlying event incorporated into them.
We also investigate a second modification that involves using
a new separation measure.  Monte Carlo studies
suggest that the standard $\kp$ algorithm performs as well as either of
these modified algorithms with respect to hadronization corrections and
reconstruction of the mass of a Higgs boson decaying to $b\bar b$.
\end{abstract}

\section{Introduction}\label{sec_intro}

Jet-finding algorithms play a central role in the study of hadronic
final states in all kinds of hard-scattering processes. Nowadays
two types of jet algorithms are in common use: the so-called
{\em cone} algorithms \cite{Snowmass,FlaMei} based ultimately
on the Sterman-Weinberg \cite{SteWei} definition in terms of
energy flow into an angular region, and the {\em clustering}
algorithms \cite{JETSET,JADE} in which jets are built up by
iterative recombination. In $\ee$ annihilation, the
so-called Durham \cite{Durham,CDOTW,BKSS} version of the
clustering algorithm has become most prevalent, although
cone algorithms have also been applied \cite{OPALcone,ChaEll}.
Up to now, cone algorithms have been more widely used in theoretical
\cite{EKS,GGK,radial} and experimental \cite{CDF,D0} studies of
jets in hadron-hadron collisions. However, clustering algorithms
have been proposed \cite{CDSW,EllSop,Seym93} and are beginning to
be applied \cite{D094,D097} in this case also. Comparative
theoretical studies \cite{Seym93,Seym94,Pump,Seym98} have
suggested that this would be beneficial, from the viewpoints
of stability with respect to higher-order and non-perturbative
corrections and improved mass resolution for heavy objects.

Recently, modifications to the Durham clustering algorithm for
$\ee$ final states have been proposed \cite{DLMW,MLS} and some
of their properties investigated \cite{MLS,BenMey}. They mainly
concern the two aspects of the algorithm which seem to be most open to
improvement. The first concerns the order in which clustering
occurs, and the second the resolution or separation measure
that is used to decide whether clustering should occur.
The general conclusion from $\ee$ studies is that a change to angular ordering can
help in the fine resolution of multi-jets and jet sub-structure,
and that a smoother separation measure can reduce sensitivity
to higher-order and non-perturbative effects. 

In the present paper we investigate the effects of analogous
modifications to the ``standard'' $\kp$-clustering
algorithm \cite{CDSW,EllSop} for hadron-hadron collisions. We
first review the criteria for a good jet algorithm, and
then, after stating the standard algorithm and the proposed
modifications precisely, we perform two types of comparative
Monte Carlo studies.

For the first set of studies, in section \ref{sec_tube},
we use a simple ``tube'' model \cite{Feyntub,hadro}, representing the hadronic final
state when no true jet production has occurred. Here the
test of a good algorithm is the extent to which it can avoid
generating spurious jets.

For the main investigations (section \ref{sec_HW}) we use the
\HW\ \cite{HW} event generator to simulate true jet production over a range of
transverse energies. We compare the jet properties at
the parton and hadron (calorimeter) levels, as well as
the sensitivity to the underlying soft event and
calorimeter threshold. We also generate Higgs boson events
and study the mass resolution for the decay $H^0\to b \bar b$.

Our conclusions are given in section \ref{sec_conc}.

\section{Criteria for a good jet-finding algorithm}\label{sec_crit}

Although a completely ``correct'' assignment of particles to jets
is not a well-defined concept, owing to quantum interference effects,
any jet algorithm must eventually assign all hadrons to a unique jet
or the ``underlying event'' for the purpose of calculations. 
Clustering algorithms naturally assign particles unambiguously,
while cone algorithms must add an {\it ad hoc} final step to
deal with hadrons in overlapping cones. Association of particles
that are close together in angle, or relative transverse momentum,
is most natural from the viewpoint of QCD dynamics, and therefore
the original JADE clustering criterion \cite{JADE} based on invariant
mass has been replaced by $\kp$-based criteria \cite{JETSET,Durham}
for the study of jets in $\ee$ annihilation.

Identification of jets in hadron-hadron collisions is harder than in $\ee$
annihilation.  In the latter case, the beams are not made of composite
particles, so the final states are cleaner.  In particular, in a
hadron-hadron collider, one must worry about ``beam jets'' and
the ``soft underlying event'' (SUE), because the
initial-state particles have quark constituents themselves.  One expects
the quarks not involved in hard scattering to continue along in a path
near the beam line, possibly emitting some low-momentum particles at
larger angles due to their soft interactions. 
The challenge is how to separate this ``beam jet'' and ``underlying''
hadronic energy from the products of the hard scattering process.
For this purpose one needs an algorithm that eliminates particles with low
transverse energy $E_T$ (relative to the beams) as early as possible from the
clustering process, if they are not close to regions of high $E_T$ activity.

At the same time any jet-finding algorithm must possess certain
basic properties in order to produce sensible (finite) results for
quantities like jet cross sections.  Included among these properties is
insensitivity to low-energy gluon radiation ({\em infrared safety}).  An
algorithm is infrared safe if it does not distinguish between a state with
an infinitely soft radiated gluon, and one without.

Also required is {\em collinear safety}.  In the case of infrared safety, a
zero energy emission can potentially lead to an infinite cross section.  A
parton emitted a zero angle can have the same effect.  To prevent this
blow-up in cross sections, the algorithm must not differentiate between two
particles emitted along the same line.  Now, it is clear that all
algorithms will trivially satisfy these criteria at the calorimeter level,
because the calorimeter will have finite resolution in energy and angle.
However, one would hope to have an algorithm that possesses these good
theoretical properties on its own. 

In the case of hadron-hadron interactions, the jet algorithm must also
allow for {\em factorization} of initial-state mass singularities.  That is,
it must allow the singularities due to initial-state collinear emissions
to be factored out and replaced by non-perturbative parton distributions,
leaving the hard scattering process to be calculated perturbatively. 

Besides the above basic requirements, there are additional criteria
one wishes to satisfy.  A good jet-finding algorithm will find a minimum
of ``spurious jets'', that is, jets formed by combining unassociated
soft particles from the underlying event or beam jets. All algorithms
will eventually give rise to such jets, at sufficiently fine resolution.
The level at which this begins to happen, and the fashion
in which the spurious jets form, is of interest. 

From an experimentalist's viewpoint, an extremely important feature for an algorithm is its ability
to reconstruct the momenta of the parent particles of the jets as
accurately as possible, for example in the hadronic decays of
heavy particles (top quarks, vector bosons, Higgs bosons, etc.).
This is an invaluable tool in new particle searches, for establishing
a potentially small signal on a potentially large
background.

Other criteria for comparison include the shifts in the values of quantities
that take place between the parton and hadron level.  It is desirable to
minimize these ``hadronization corrections''.

In this paper, we compare
spurious jet formation, hadronization effects, the ability to pick
jets out of the underlying event, and the reconstruction of invariant
masses, for the standard $\kp$-clustering algorithm and two possible
variants of it.

\section{Algorithm definitions}\label{sec_algdefs}

We give here the precise definition of the standard $\kp$-clustering
algorithm \cite{CDSW,EllSop} and then the proposed new variants.
The choice of variables and combination procedure differ
somewhat from those used for $\ee$ physics. This is primarily because
the hard parton scattering process has an unknown longitudinal boost
in hadron-hadron collisions. One therefore needs to use variables
that are approximately invariant under such boosts.

First we require some general definitions, as follows:
\begin{itemize}
\item Define a dimensionless {\it cone size} parameter $R$
(usually $R\leq 1$).

\item For each object $i$ (hadron, calorimeter cell or whatever) in the
final state, define the azimuthal angle $\phi_i$, the pseudorapidity
\beq  \eta_i = -\ln\tan\left(\frac{\theta_i}{2}\right) 
\eeq
and the transverse energy
\beq
E_{Ti} = E_i\sin\theta_i 
\eeq
where $\theta_i$ is the polar angle with respect to the direction
of one of the incoming beams in the overall c.m. frame.

\item The {\it separation from the beams} of object $i$ is
\beq\label{didef} d_i=E_{Ti} R\;. \eeq

\item For each pair of objects $i$ and $j$ the {\em angle} is
\beq R_{ij} = \sqrt{(\eta_i-\eta_j)^2+(\phi_i-\phi_j)^2} 
\eeq
and the {\it separation} is
\beq d_{ij} = \min\{E_{Ti},E_{Tj}\} R_{ij}\;. 
\eeq

\item The {\it combination procedure} $i+j \to (ij)$ is
\beeq\label{combdef}
E_{Tij} &=& E_{Ti} + E_{Tj} \nonumber \\
\eta_{ij} &=& \frac{(E_{Ti}\eta_i + E_{Tj}\eta_j)}{E_{Tij}} \\
\phi_{ij} &=& \frac{(E_{Ti}\phi_i + E_{Tj}\phi_j)}{E_{Tij}} \nonumber
\eeeq
\end{itemize}

\subsection{\boldmath Standard $\kp$ algorithm}
We use here the algorithm proposed in ref.~\cite{CDSW} with
the modifications suggested in ref.~\cite{EllSop}.
\begin{enumerate}

\item Find the pair $I$ and $J$ with $d_{IJ}=\min\{ d_{ij}\}$.

To determine whether clustering occurs, we compare this separation with
the separation of the object ``nearest'' the beam.

\item Find the object $K$ with $d_K=\min\{ d_i\}$.

\item If $d_{IJ}\leq d_K$ then combine  $I+J \to (IJ)$, update
the lists of $d_{ij}$'s and $d_i$'s accordingly, and go to 1.

\item If $d_{IJ} > d_K$ then store $K$ as a {\it completed jet},
remove it from the lists of $d_{ij}$'s and $d_i$'s, and go to 1.

\item Iterate until all jets are completed.

\end{enumerate}
The presence of the cone size $R$ in Eq.~\re{didef} ensures that objects
cannot be clustered unless their separation is less than $R$, since
otherwise either $d_I$ or $d_J$ is less than $d_{IJ}$ and step four
removes one of them.

Step four is also roughly akin to the ``soft-freezing'' that is used in the
recently proposed Cambridge $\ee$ algorithm \cite{DLMW}. By removing the less
energetic objects from the list as clustering occurs, we prevent them
from attracting additional partners that could build spurious jets.

\subsection{\boldmath Proposed improved algorithm: $AO_{1}$}
The $AO_{1}$ algorithm is an angular-ordered version of the standard
algorithm. The basic idea of an angular-ordered algorithm
is to use the angular variable $R_{ij}$, instead of the separation $d_{ij}$,
to choose the pair of objects to be considered first for clustering.

\begin{enumerate}

\item Find the pair $I$ and $J$ with $R_{IJ}=\min\{ R_{ij}\}$.

\item Find the object $K$ with $d_K=\min\{ d_i\}$.

\item If $d_{IJ}\leq d_K$ then combine  $I+J \to (IJ)$, update
the lists of $R_{ij}$'s and $d_i$'s accordingly, and go to 1.

\item If $d_{IJ}> d_K$, then we must check whether the particle pair
still remains within the angular cone size, $R$. If it does ($R_{IJ}\leq R$),
then move to the pair
$I'$ and $J'$ with the next-smallest value of $R_{ij}$ and repeat
step 3, with $I'$ and $J'$ in place of $I$ and $J$. 

\item If the particle pair does not lie within the fixed cone size
($R_{IJ} > R$) then store $K$ as a {\it completed jet},
remove it from the lists of $R_{ij}$'s and $d_i$'s, and go to 1.

\item Iterate until all jets are completed.

\end{enumerate}

\subsection{\boldmath Angular-ordered algorithm with new separation
measure: $AO_{2}$}
For the second angular-ordered algorithm we use the same steps as in
$AO_1$ but change the definition of the separation measure between any
two objects to
\beq
d_{ij}'
= \frac{2E_{Ti}E_{Tj}}{(E_{Ti}+E_{Tj})}R_{ij} \geq d_{ij}\;.\eeq 
The separation between an object and the beam remains as before
\beq d_{i}'=E_{Ti}R = d_i\;.\eeq

\begin{figure}
\epsfbox[40 540 550 770]{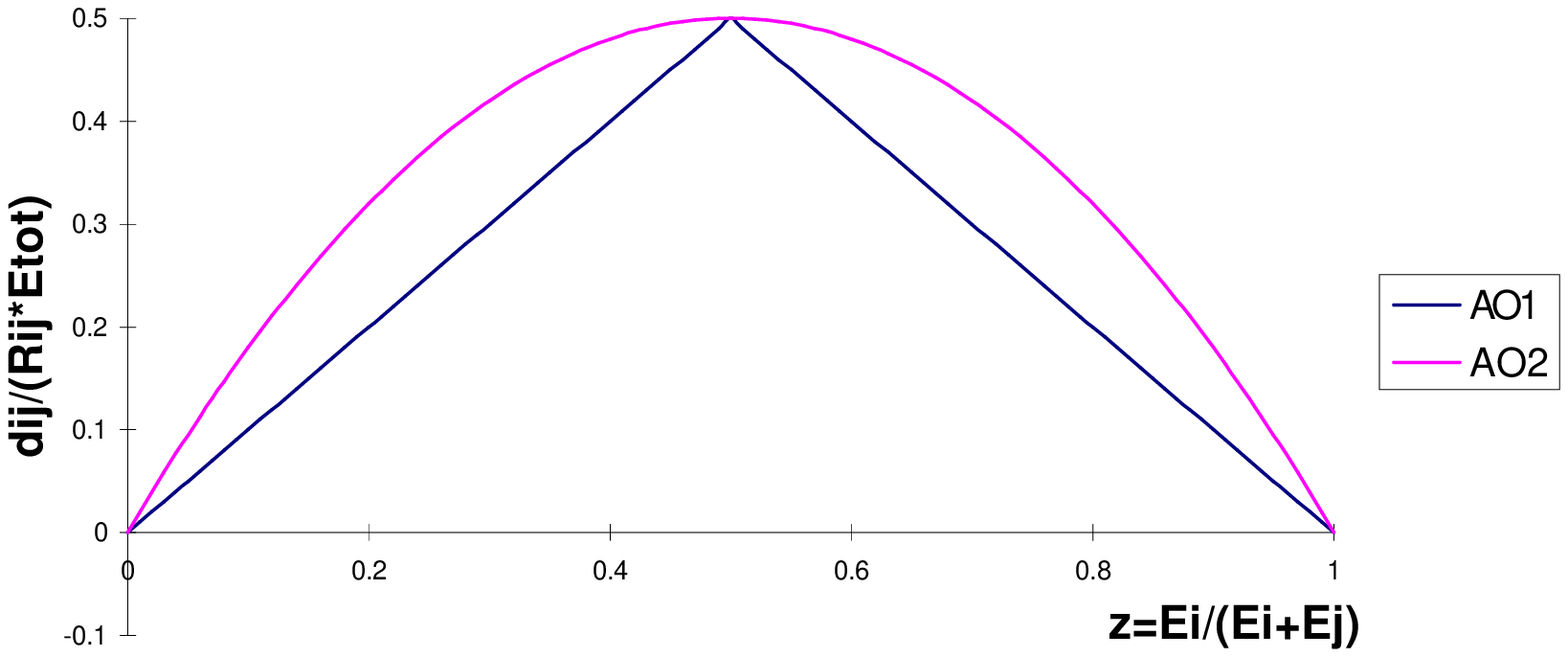}
\caption{A comparison of different separation measures.  We fix the total
energy $E_{TOT}=E_{i}+E_{j}$ and vary
$z=\frac{E_{i}}{E_{i}+E_{j}}$ from 0 to 1.}\label{fig_sep}
\end{figure}

This $d_{ij}'$ measure has interesting properties.  It retains the same
basic shape as the original definition of $d_{ij}$, but it is now a
smoothly varying function.  This is shown in figure \ref{fig_sep}.
It is analogous to
the {\tt LUCLUS} separation measure \cite{JETSET} used in $\ee$
physics, where it has been found to have good features \cite{MLS}.

\vspace{.3in}

\section{Theoretical properties of \mbox{algorithms}}\label{sec_theo}
It is immediately clear that all the above algorithms satisfy
the infrared safety
criterion.  Recall that infrared safety is obtained if measured jet
variables are unchanged when an $E_{T}\to 0$ parton is emitted.  From
the definition of $d_{ij}$ (or $d_{ij}'$), it is clear that the proposed
algorithms will immediately combine such a gluon with the hard parton that
emitted it.

Similarly, it is clear that the algorithms are collinear safe.  If a
parton emits another parton along the same direction, the algorithms will
immediately combine the two. Therefore, there will be no difference in the
observable quantities in the case of collinear emission. 

The algorithms also allow factorization of initial state
singularities.  As $ \theta_i\to 0$, the separation from the beam,
$d_i\to 0$.  Therefore, the beam jets do not affect the
combination of other jets, as hadrons collinear to the beams are
immediately classified as completed jets by the algorithm,
and taken off the combination list. This allows the
remaining hadrons to go about combination unhindered. 

\begin{figure}
\epsfbox{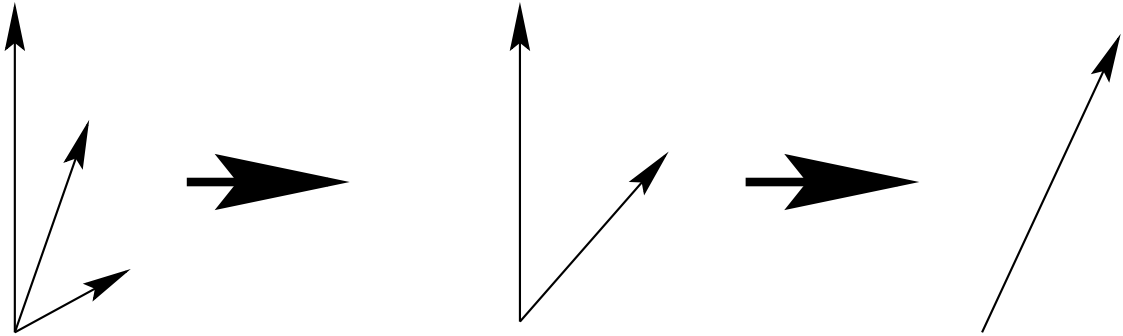}
\caption{A misclustering of jets by the standard $\kp$-clustering algorithm.}
\label{fig_misc}
\end{figure}

Angular-ordered algorithms attempt more possible clusterings than
the standard algorithm, because they may consider other pairs of objects
before coming to the pair with the smallest separation under the $d_{ij}$
metric. However, any tendency towards over-clustering is balanced by the
fact that the
algorithms tend to start at the middle of each jet.  As a consequence,
we would hope that they tend to pick up fewer unassociated particles on
the periphery of the jet.  A rough schematic of such a case is
shown in figure \ref{fig_misc}. 
In the standard algorithm, the two low-energy particles may
combine. Then because the axis of the two has been
pulled toward the centre, this composite jet may combine with the third
particle.  In the angular-ordered case, the jet is built up from the
centre.  Thus, the two central particles can combine, leaving the wide-angle
particle to be resolved separately.

Cases such as these would lead us to hope that the angular-ordered
algorithms might be able to separate the interesting high-$E_{T}$ jets more
effectively from the soft underlying event.  However, the above schematic
only works in a fairly limited kinematic window.  In most cases, the
wide-angle radiation will still lie within the cone size $R$, and combination
will still take place as in the standard $\kp$-algorithm.

\section{Tube model studies}\label{sec_tube}
Here a ``tube'' of particles is taken to be a cylinder of particles
uniformly distributed in $\phi$ and $\eta$.  The transverse momenta
of the particles in the tube have a strongly damped exponential
distribution.  This simulates the hadronization of a soft hadronic
collision or underlying event. There are no true jets, so any jet
found is spurious, which is a useful comparative test of
algorithms \cite{DLMW}.

The particle multiplicity in this model is given by
\beq
\VEV{n} = 2\frac{\mu}{\VEV{p_t}}
\log\left(\frac{E_{\mbox{\tiny CM}}}{\mu}\right)
\eeq
where $\mu$ is the invariant mass per unit rapidity (here set to 0.5 GeV),
$E_{\mbox{\tiny CM}}$ is the centre-of-mass energy, and $\VEV{p_t}$ gives
the average transverse momentum per particle (again set to 0.5 GeV).
The model was run at centre-of-mass energy 1 TeV, which gave
$\VEV{n}\simeq 15$ particles per event.  Each algorithm was run on
10,000 events.

It was found that there was no difference in jet formation between 
the standard $\kp$ algorithm and the $AO_{1}$ algorithm. 
We looked at the number of particles included in the jet
with the maximum $E_T$ (see table \ref{tab_mult}).  It is expected 
that the number of particles in this jet would be 
indicative of the amount of clustering that is done
by the algorithm, as it is likely that if particles were combined, they would
combine into a jet with a higher $E_{T}$ than that of a single particle.

\begin{figure}
\epsfbox[45 525 555 780]{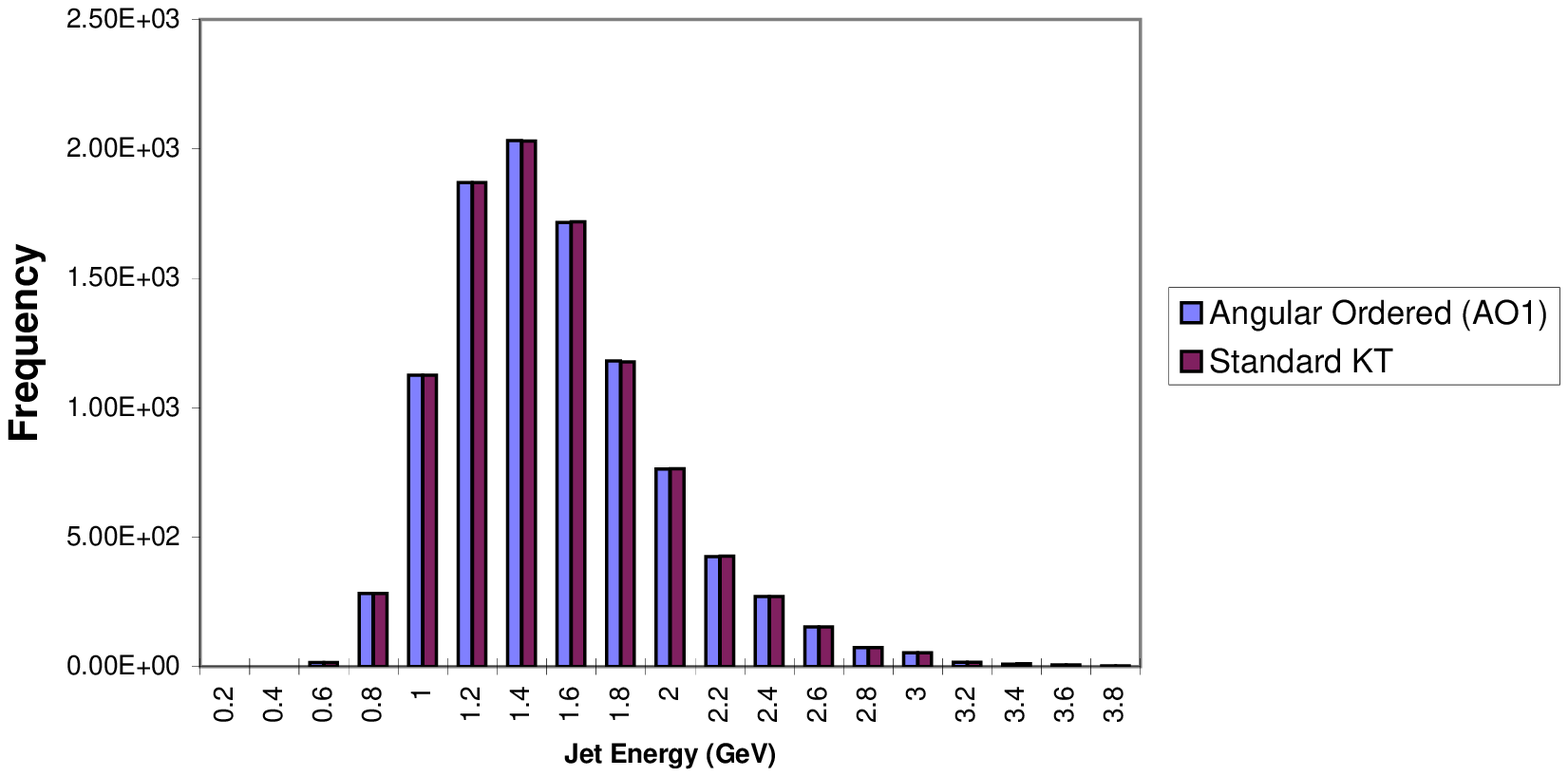}
\caption{$E_{T}$ spectra of maximum $E_{T}$ jet in the tube model.}
\label{fig_tubeao1}
\end{figure}

We notice a striking similarity of $AO_{1}$ to the standard $\kp$-clustering
algorithm by examining the energy spectra of the maximum $E_{T}$ jets 
(see figure \ref{fig_tubeao1}).
If additional spurious jet formation occurred in
either case, one would expect to find that energy spectra to be broader
and shifted to a higher energy. However, this signature is clearly absent,
and the spectra are identical.

\begin{figure}
\epsfbox[50 535 555 770]{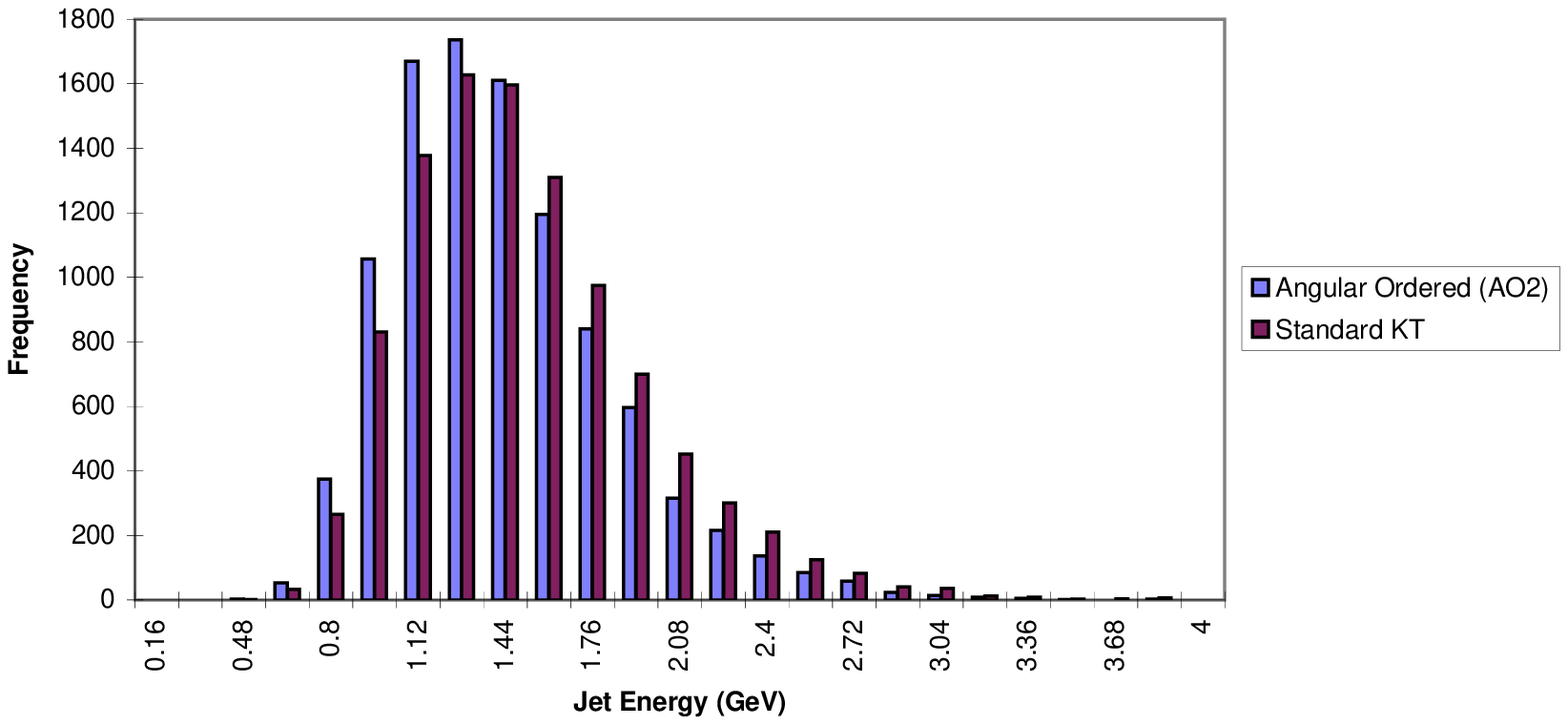}
\caption{$E_{T}$ spectra of maximum $E_{T}$ jet in the tube model.}
\label{fig_tubeao2}
\end{figure}

On the other hand, there is indeed a difference between the
standard $\kp$ algorithm and the $AO_2$ algorithm (figure \ref{fig_tubeao2}). 
Less clustering occurs in the case of the $AO_{2}$
algorithm.  This is a result of the definition of the separation measure,
$d_{ij}'$.  Notice that in figure \ref{fig_sep}
the separation measure used in $AO_2$
is greater than that for $AO_1$, for all values of $E_{T}$.  As a
result, the clustering condition will be satisfied less
frequently. Consequently, the energy spectrum for the new
separation measure $d_{ij}'$ is shifted to lower energies,
indicative of less clustering.

\begin{figure}
\epsfbox[45 525 540 770]{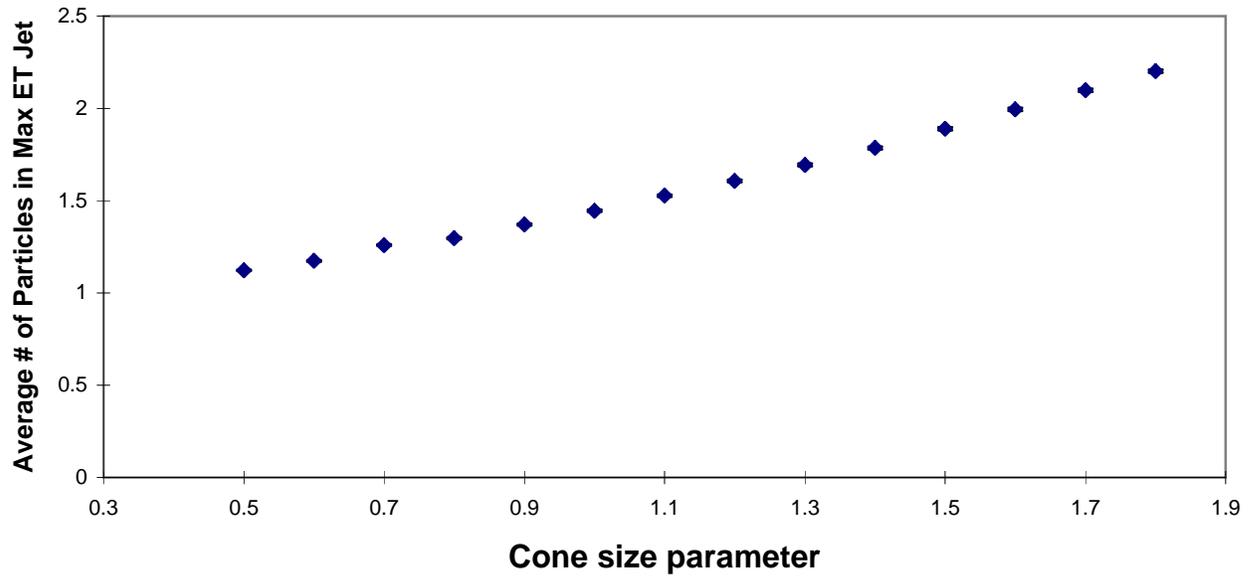}
\caption{A measure of the amount of clustering done by the $AO_{2}$
algorithm as the cone size parameter is varied.}
\label{fig_tubeR}
\end{figure}

It is interesting to vary the cone size parameter, $R$, to see where a
comparable amount of clustering occurs.  As seen in
figure \ref{fig_tubeR}, it is
found that this happens when $R$ is increased from unity
to a value of approximately $R=1.3$.

\section{HERWIG Monte Carlo \mbox {studies}}\label{sec_HW}

\HW\ \cite{HW} is a Monte Carlo event generator that can be used to
simulate many kinds of events.  First the hard scattering process is
calculated perturbatively, and then parton showers are generated
until a given scale is reached, on the order of the hadron
masses.  Hadronization is simulated by combining
partons into colorless clusters, which decay to hadrons.  The model
conserves local flavour and energy momentum flow.  The underlying event is
modelled as a separate collision between colorless clusters containing
hadron remnants. For our study, QCD events were generated in
symmetric $p\overline{p}$ collisions at a centre-of-mass energy
of 1.8 TeV, as at Tevatron Run I.

For each type of test conducted, we attempted to ascertain the difference
between applying a jet algorithm at the parton and calorimeter level. 
In order
to run the algorithms at the parton level, the \HW\ simulation was
simply stopped before hadronization, and the clustering was conducted on
the partons at this stage.  This is not entirely equivalent to the
calculation of quantities via perturbation theory, but should still
give a good comparative indication of hadronization effects.
At the calorimeter level, a simple segmentation
simulation was used \cite{CALO}. The calorimeter was segmented into
63 equal cells in $\phi$, for a
resolution of $.1$ radians and 100 cells in pseudo-rapidity covering a
region from $-5\leq\eta\leq5$, for the same rapidity resolution.  The
energy resolution of each cell is given by: 
\beq
\frac{\sigma(E)}{E}=\frac{{\tt RES}}{\sqrt{E}}
\eeq
where {\tt RES} is a constant that differs for the hadronic and
electromagnetic calorimeters.  As a rough guide, we used ${\tt RES}=0.1$
for the electromagnetic and ${\tt RES}=0.5$ for the hadronic calorimeters
($E$ being measured in GeV).

To cover a range of jet energies, we varied the \HW\ parameter {\tt PTMIN},
which sets the minimum parton transverse energy for the hard scattering
processes.  By varying this parameter, we examined jet transverse
energies from 30 to 90 GeV with comparable statistics. A thousand
events were collected at each value of {\tt PTMIN}.

\subsection{Shift in jet axis}
For each algorithm, we investigate the shift in jet axis, $\Delta R$, given by:
\beq
\Delta R=\sqrt{
 (\eta_{\mbox{\tiny parton}}-\eta_{\mbox{\tiny calorimeter}})^{2}
+(\phi_{\mbox{\tiny parton}}-\phi_{\mbox{\tiny calorimeter}})^{2}}
\eeq
This measure gives an indication of whether the algorithm is
doing a good job of
preserving the direction of parton energy flow at the calorimeter level.
Depending on how invariant masses are reconstructed, this can have a
radical effect on the masses of reconstructed particles.

Comparisons of the shift in the jet axis between parton level and
calor\-imeter level show little difference between the algorithms.
The scheme used to pair the calorimeter jets with parton jets was
rudimentary.  It simply took the two highest-$E_T$ jets and matched them
in such a way as to minimize the total $\Delta R$ for these two jets.  
Other schemes are possible, such as optimizing over a larger number of
jets, or looking for the highest-$E_T$ jets on opposite sides in azimuth.
Variations of the $\Delta R$ matching scheme are a possible topic for
future investigation. As shown in table \ref{tab_shift}, the values of
$\Delta R$ agree within
statistical error.  This is a somewhat surprising result, given the
angular nature of the new algorithm. 

Note also that the calorimeter has finite angular resolution,
$\Delta R\simeq 0.1$.  This serves to increase the value of $\Delta R$
for each algorithm, but should not affect any one more than
the others.  It seems, then, that all three algorithms that were
investigated have similar shifts in jet axes.

\subsection{Jet radial moment}\label{sec_radial}
We also examine the change in the jet radial moment, which is a measure of how
the energy is distributed about the jet axis.  The radial moment is defined
as \cite{radial,Seym98}:
\beq\label{jetrad}
\VEV{R}=
\sum_{\mbox{\tiny constituents}} \frac {E_{Ti}R_{i}}{E_{\mbox{\tiny jet}}}
\eeq
where $R_{i}$  is the angular separation of the $i$th parton or
calorimeter cell from the jet axis.  A more compact jet, with its
energy concentrated near
the jet axis, has a smaller value of $\VEV{R}$.

\begin{figure}
\epsfbox[50 540 545 775]{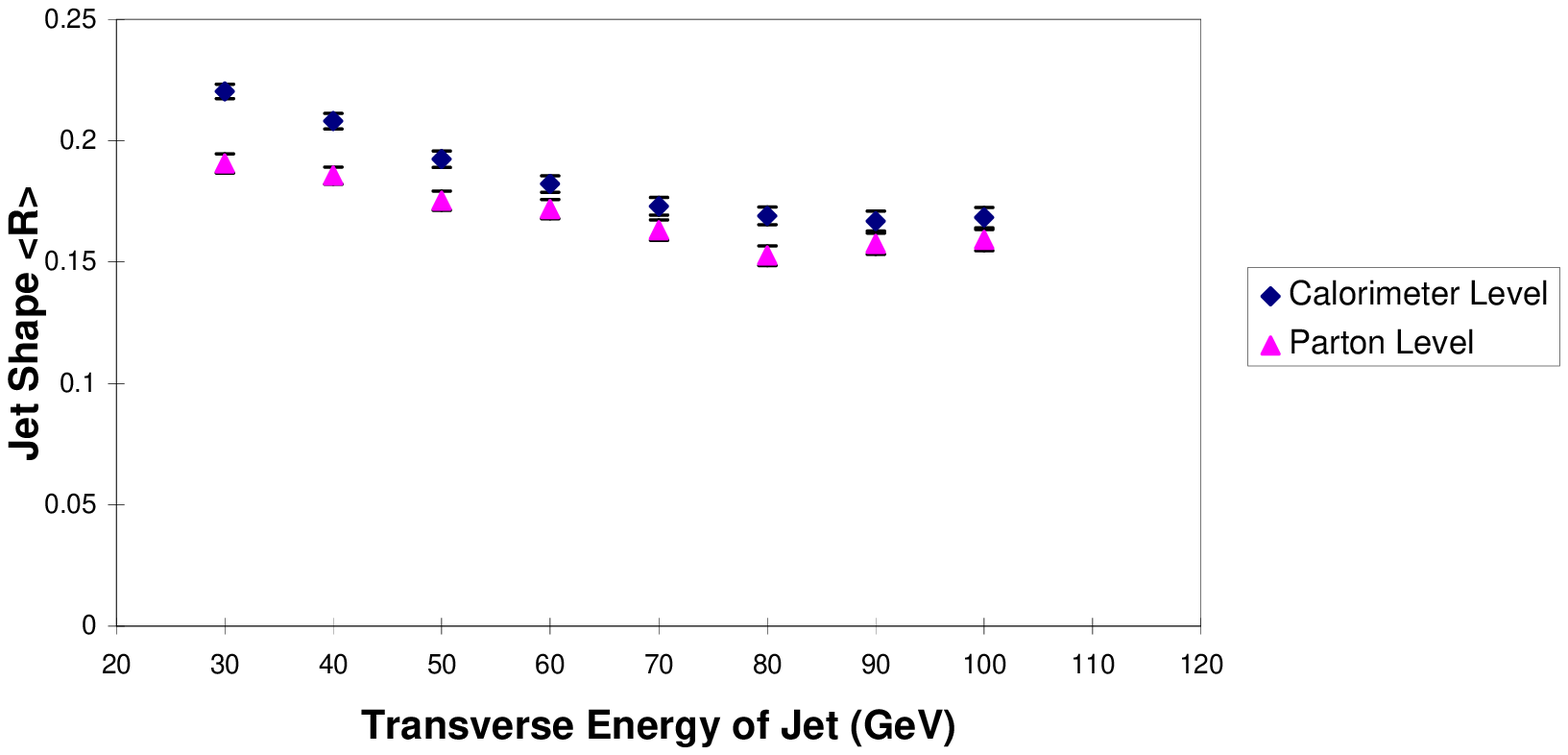}
\vspace{.2in}
\epsfbox[50 540 545 775]{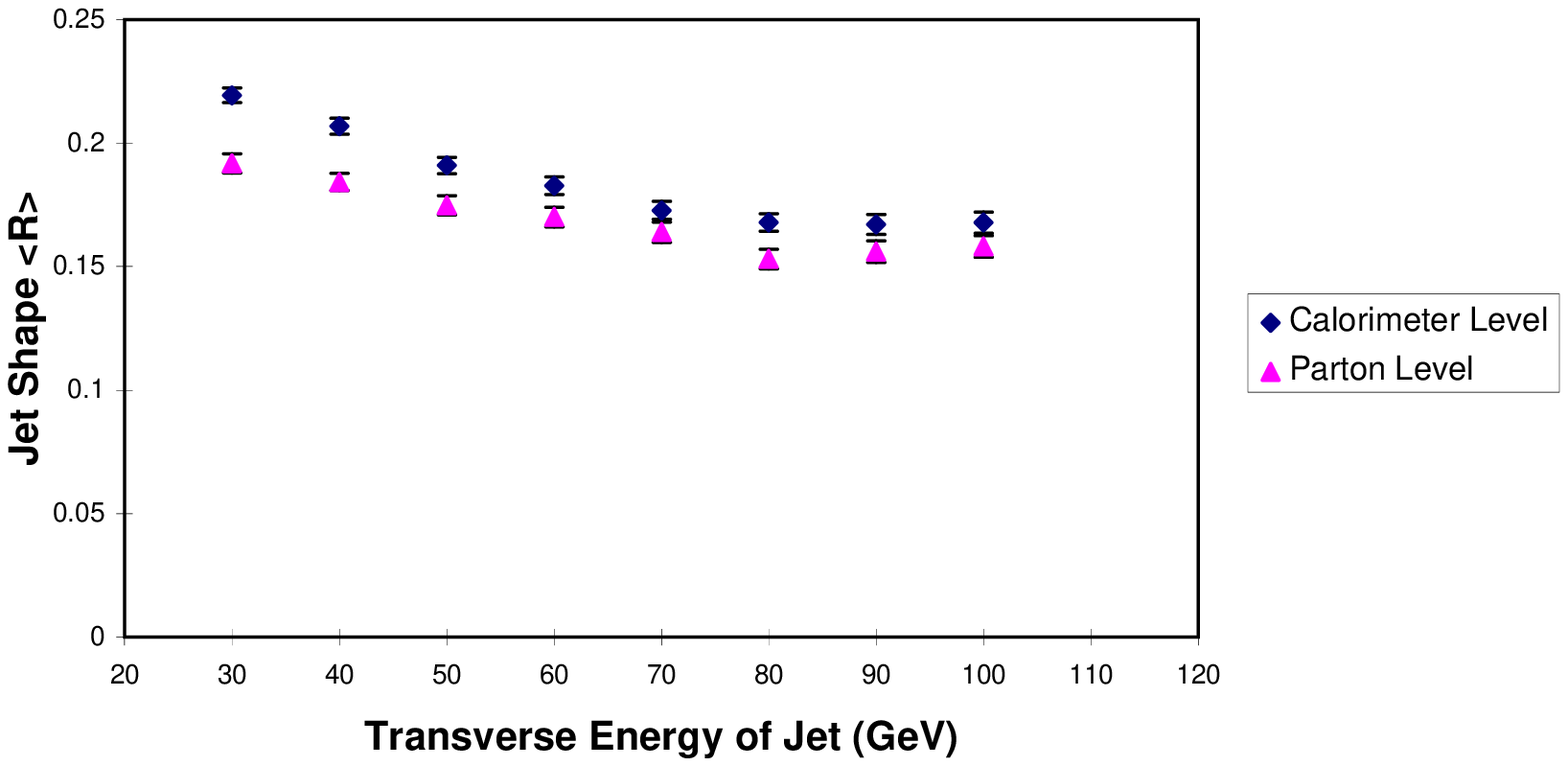}
\vspace{.2in}
\epsfbox[50 530 530 775]{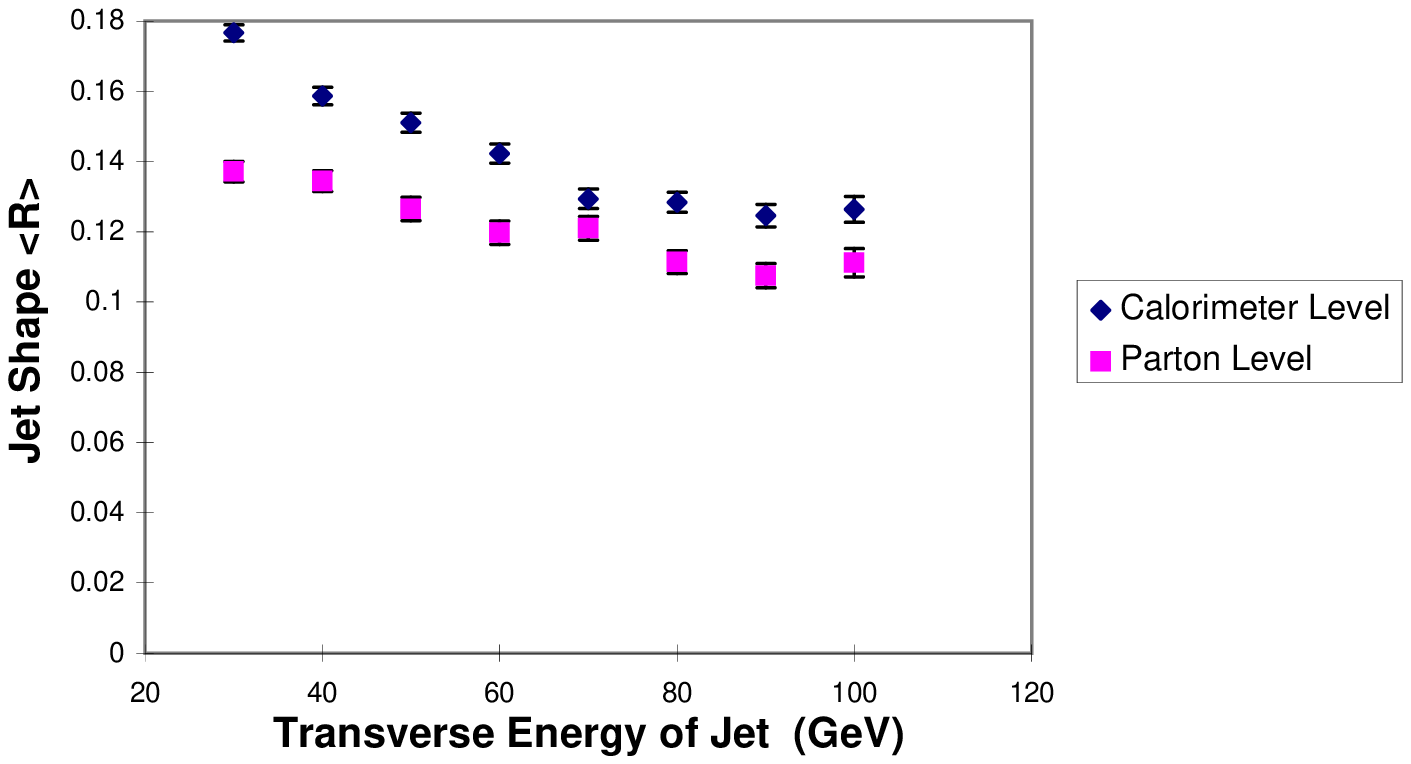}
\caption{Radial moment curves for (a) Standard $\kp$, (b) $AO_{1}$ and
(c) $AO_{2}$. The curves are constructed from 1000 events at each of
{\tt PTMIN}=30,50,70,90.}
\label{fig_shapes}
\end{figure}

\begin{figure}
\epsfbox[50 535 535 770]{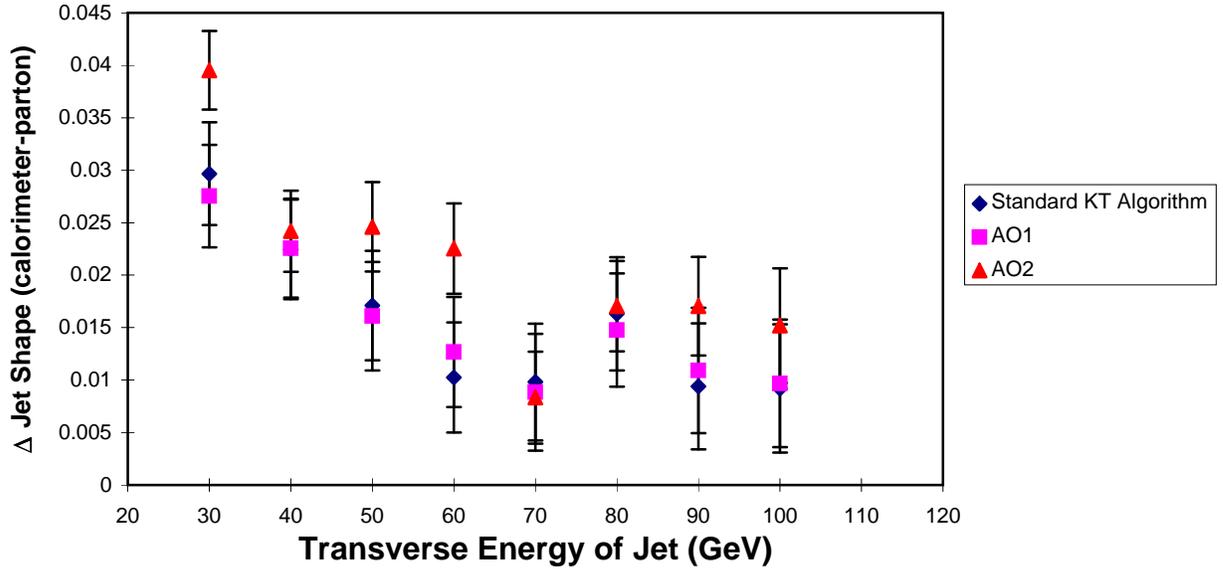}
\caption{Difference between the calculated jet shapes at
the parton and calorimeter level for each of the three algorithms
under investigation.  Each curve here is obtained by taking the difference
between the two curves in each of the figures in figure
\ref{fig_shapes}.}
\label{fig_delsh}
\end{figure}

In a comparison between the radial moments, as defined in
equation \re{jetrad}, it seems that all of the algorithms
have similar discrepancies between the parton and
calorimeter levels (see figure \ref{fig_delsh}). 
However, note that the radial moment for
the $AO_{2}$ algorithm is smaller.  Again, this more collimated jet is a
result of the larger separation measure (see figure \ref{fig_shapes}).

As a consequence of these results on the radial moment,
one would expect that the standard
$\kp$-clustering algorithm and the $AO_{1}$ algorithm would perform similarly
in the reconstruction of Higgs masses, while $AO_{2}$ would, in general,
reconstruct lower masses. In section \ref{sec_higgs}, we find
that the algorithms do, in fact, follow this behaviour.

\subsection{Jet energy spectra}
The single jet inclusive transverse energy spectra were also examined
for each of the algorithms.  If more combination of
particles took place for a given algorithm, one would expect to see a rise in cross section for
higher jet energies, and a corresponding decrease for
lower jet energies.

\begin{figure}
\epsfbox[50 535 535 770]{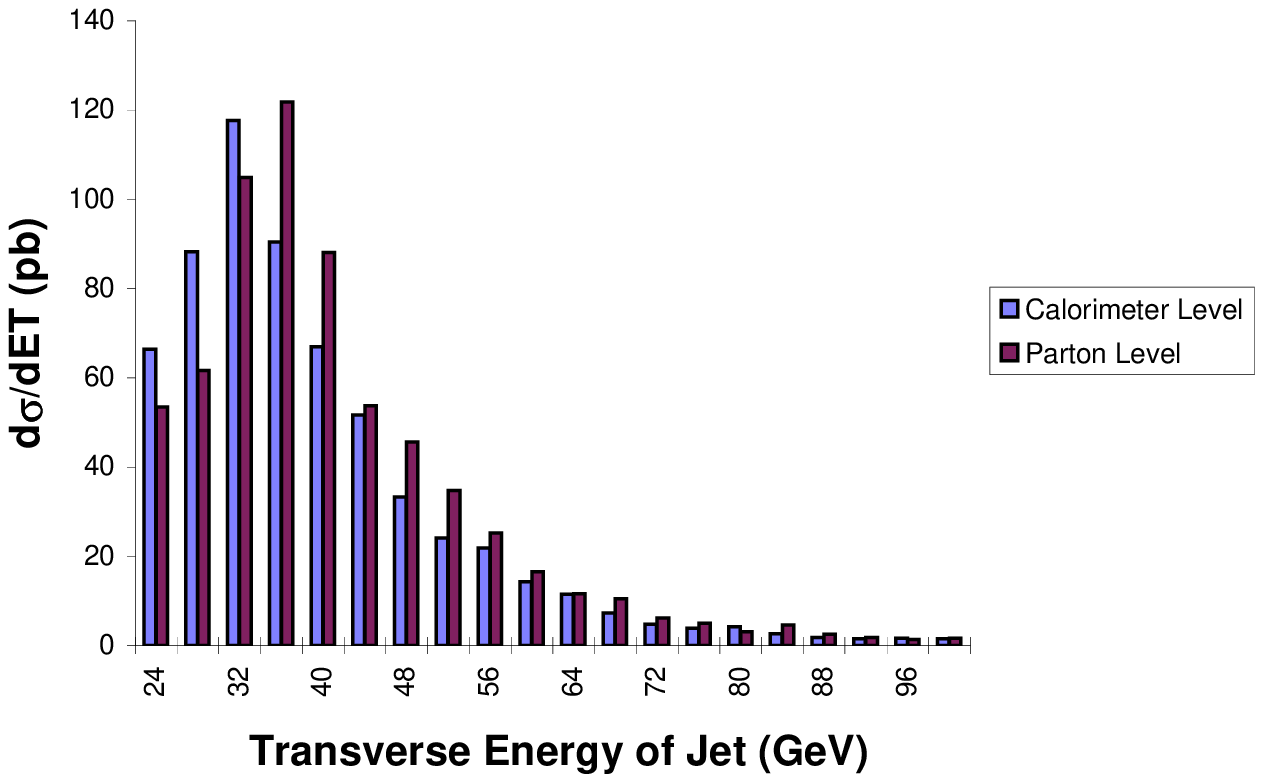}
\vspace{.2in}
\epsfbox[50 535 535 770]{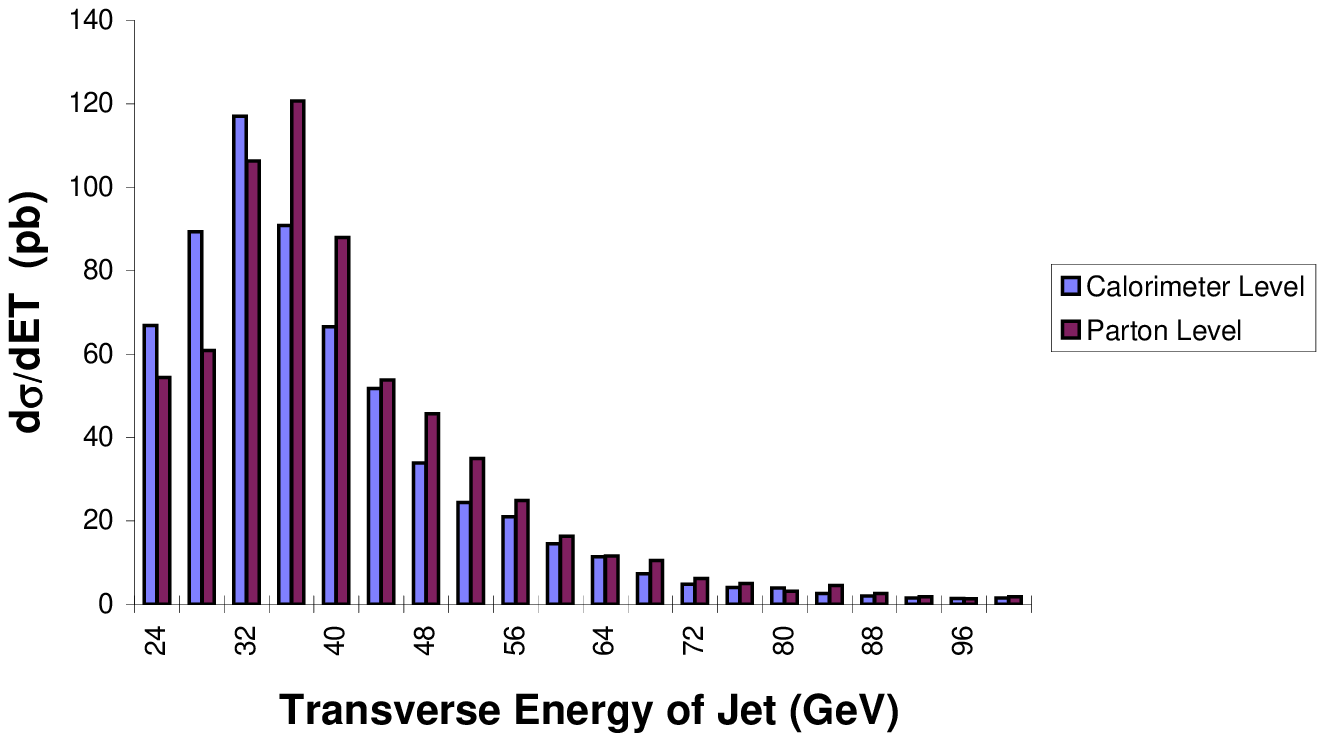}
\vspace{.2in}
\epsfbox[50 535 535 770]{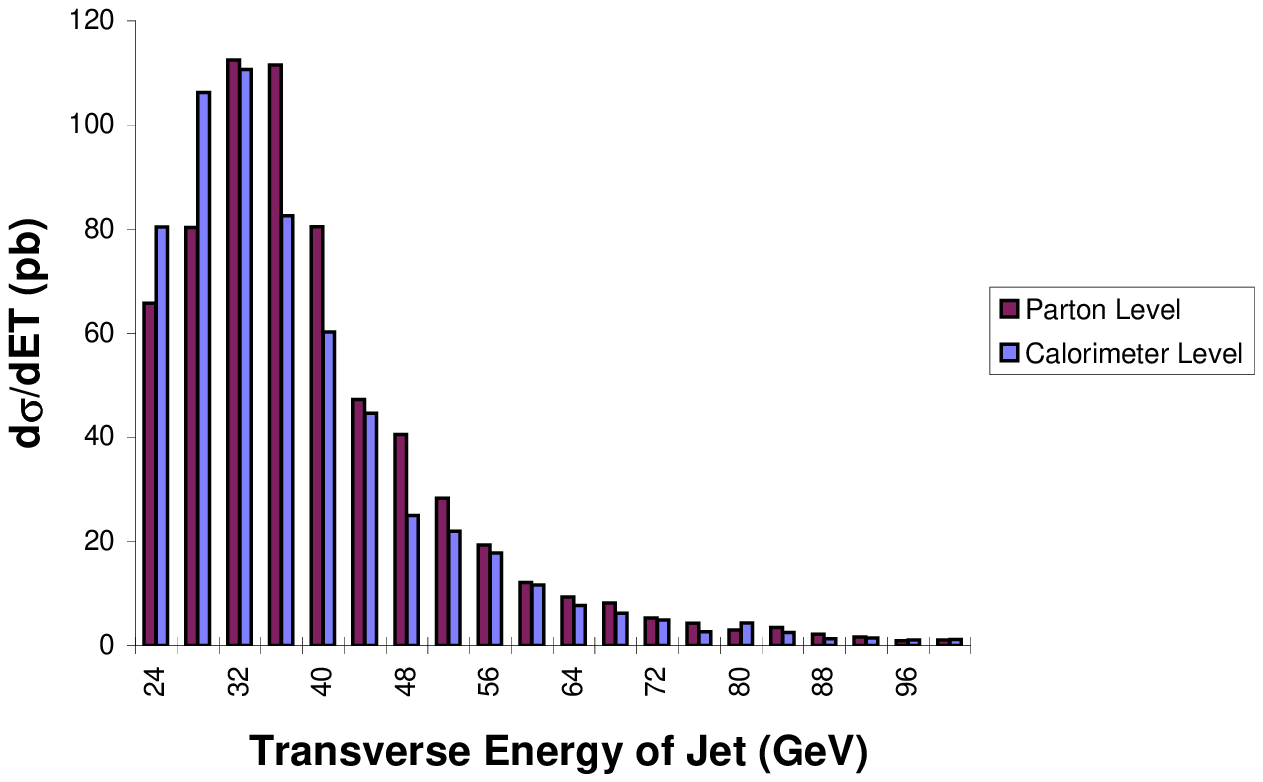}
\caption{Transverse energy spectra of (a) Standard $\kp$, (b) $AO_{1}$,
and (c) $AO_{2}$.}
\label{fig_spectra}
\end{figure}

The spectra for the three algorithms are shown in figure \ref{fig_spectra}.
The standard $\kp$ algorithm and the $AO_{1}$ are quite similar to one
another. Moreover, the results at the calorimeter and parton level are similar.
This indicates relatively small hadronization corrections.  Note, however,
that both algorithms show a tendency to push the calorimeter spectrum to lower
energies. This is most likely a consequence of forming spurious
low-energy jets. 

The spectrum shows an overall shift to lower energies in the case of the
$AO_{2}$ algorithm. This is expected on account of the lower effective jet
radius $R$ (see section \ref{sec_tube}).
It also seems to show slightly less distortion of the spectrum
between the parton and the calorimeter levels. 

\subsection{Sensitivity to calorimeter threshold energy}

It was also investigated whether the two algorithms differed in their
sensitivity to the threshold energy in the calorimeter cells.  In the
radial moment analysis (section \ref{sec_radial}) only cells having 
energies greater than 1 GeV were included as initial ``particles'' for
the algorithms to cluster.  

This reduces the multiplicity of final-state particles to a
reasonable number, thereby speeding up computing.  The threshold value was
changed from 1 GeV to 0.5 GeV, and 1000 events were generated at a minimum
hard process scale ({\tt PTMIN}) of 50 GeV.

\begin{figure}
\epsfbox[50 535 535 770]{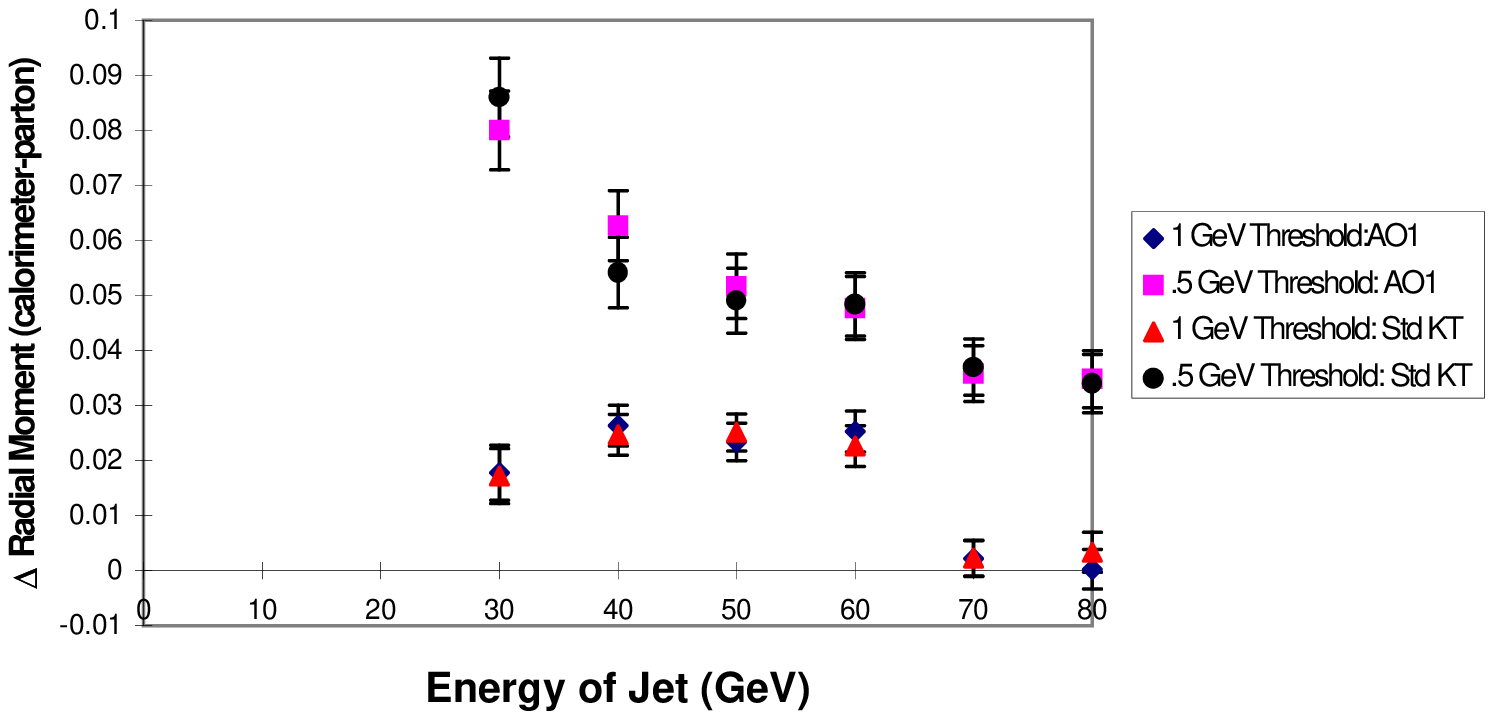}
\caption{A comparison between the standard $\kp$ and $AO_{1}$ algorithms at different
calorimeter thresholds.  Here we examine the difference in the radial moment
between the calorimeter and the parton levels.}
\label{fig_thresh1}
\end{figure}

Indeed all algorithms did show some sensitivity to the threshold.  As expected,
the radial moment increased at the smaller threshold.  This corresponds to
an increased number of calorimeter cells in the jet.  However, the $AO_{1}$
algorithm and the standard $\kp$-clustering algorithm seem to be
equally sensitive
to the change in threshold energy, as shown in figure \ref{fig_thresh1}.
The $AO_{2}$ algorithm seems to be slightly less sensitive to variation
in the calorimeter threshold (figure \ref{fig_thresh1}).

All three algorithms displayed similar behaviour with respect to changes in
shifts in the jet axis when the threshold was varied.  This is illustrated
in table \ref{tab_calshift}. All three algorithms seem to perform in an
identical manner in this test case.

\subsection{Sensitivity to soft underlying events (SUE)}
Because of an angular-ordered algorithm's tendency to incorporate fewer
particles at wide angles to the jet, one would hope that it would be less
sensitive to the soft underlying events (SUE).  

\begin{figure}
\epsfbox[50 535 535 770]{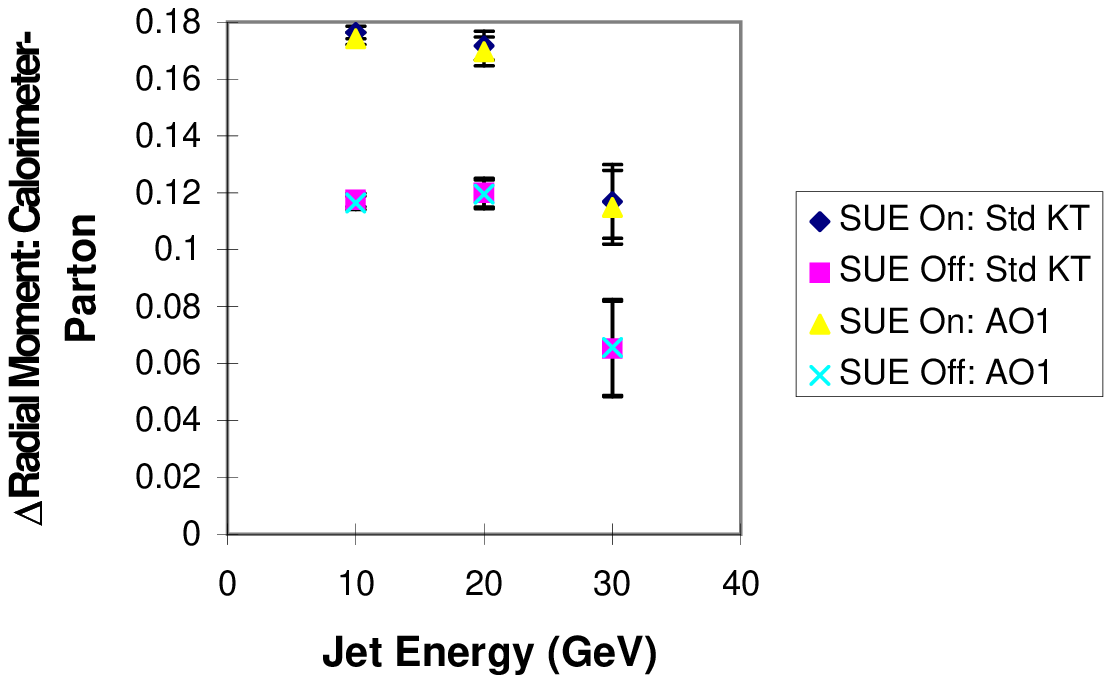}
\caption{A comparison between the standard $\kp$ and $AO_{1}$ algorithms with and
without soft underlying events.  The plot is generated from 1000 events at
a minimum hard scattering energy ({\tt PTMIN}) of 10 GeV.  The calorimeter
threshold energy is 0.4 GeV.}
\label{fig_sue1}
\end{figure}

\begin{figure}
\epsfbox[50 535 535 770]{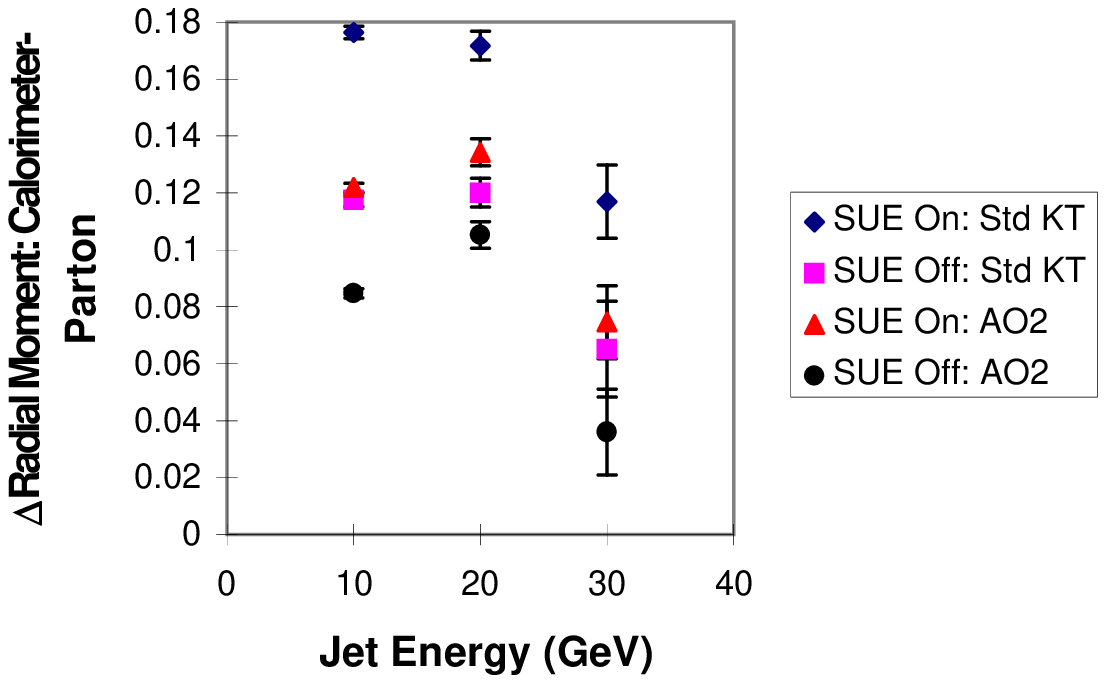}
\caption{A comparison between the standard $\kp$ and $AO_{2}$ algorithms with and
without soft underlying events.  The plot is generated from 1000 events at
a minimum hard scattering energy ({\tt PTMIN}) of 10 GeV.  The calorimeter
threshold energy is 0.4 GeV.}
\label{fig_sue2}
\end{figure}

In \HW, one can turn the underlying event on and off with a
simple switch.  Runs were made with and without the {SUE} at a lower
{\tt PTMIN}=10 GeV.  With the minimum hard scattering scale at a lower energy,
one would expect the algorithms to have a difficult time separating the
high-$E_T$ jets from the {SUE}.  Indeed, we find the hadronization
effects for all algorithms are worse with the {SUE}.
However, the algorithms seem to display nearly identical reactions to the
presence of a {SUE}. We find that the hadronization corrections
are not reduced for either
angular-ordered algorithm (figures \ref{fig_sue1} and \ref{fig_sue2}).

\subsection{Reconstruction of Higgs mass}\label{sec_higgs}
We also used \HW\ to generate Monte Carlo events containing a 
Standard Model Higgs boson with a mass of 110 GeV. At this mass, the
Higgs boson decays almost entirely through the channel
\beq H^0\to b \overline{b} \eeq 
Events were generated in symmetric $p\overline{p}$ collisions at a
centre-of-mass energy of 2 TeV, similar to the energy at Tevatron Run II.

In order to reconstruct the Higgs mass, one must select the two jets from
which the Higgs boson will be reconstructed. We reconstructed the Higgs mass
at the calorimeter level by simply looking at the invariant mass of the two
most energetic jets.  In an actual experiment, $b$ vertexing information could
be taken into account to help pick the correct jets.  Our method of jet
selection is too simplistic, but it should be sufficient for comparative
purposes.

The first feature to notice is that all the algorithms reconstruct the Higgs
mass with a large width.  The natural width of the Higgs boson
at this energy is about 50 MeV, whereas the spread in the reconstructed mass
is on the order of 30 GeV.

In the case of a calorimeter threshold of 1 GeV, every algorithm examined
tends to undershoot the Higgs mass.  This is partially due to resolving
part of the $b$-jets into separate jets, which are not
included in the final mass.  In addition, energy can be lost to calorimeter
cells that do not reach the threshold.  Therefore, one would expect that the
reconstructed mass would depend on the calorimeter threshold.  

\begin{figure}
\epsfbox[50 535 535 770]{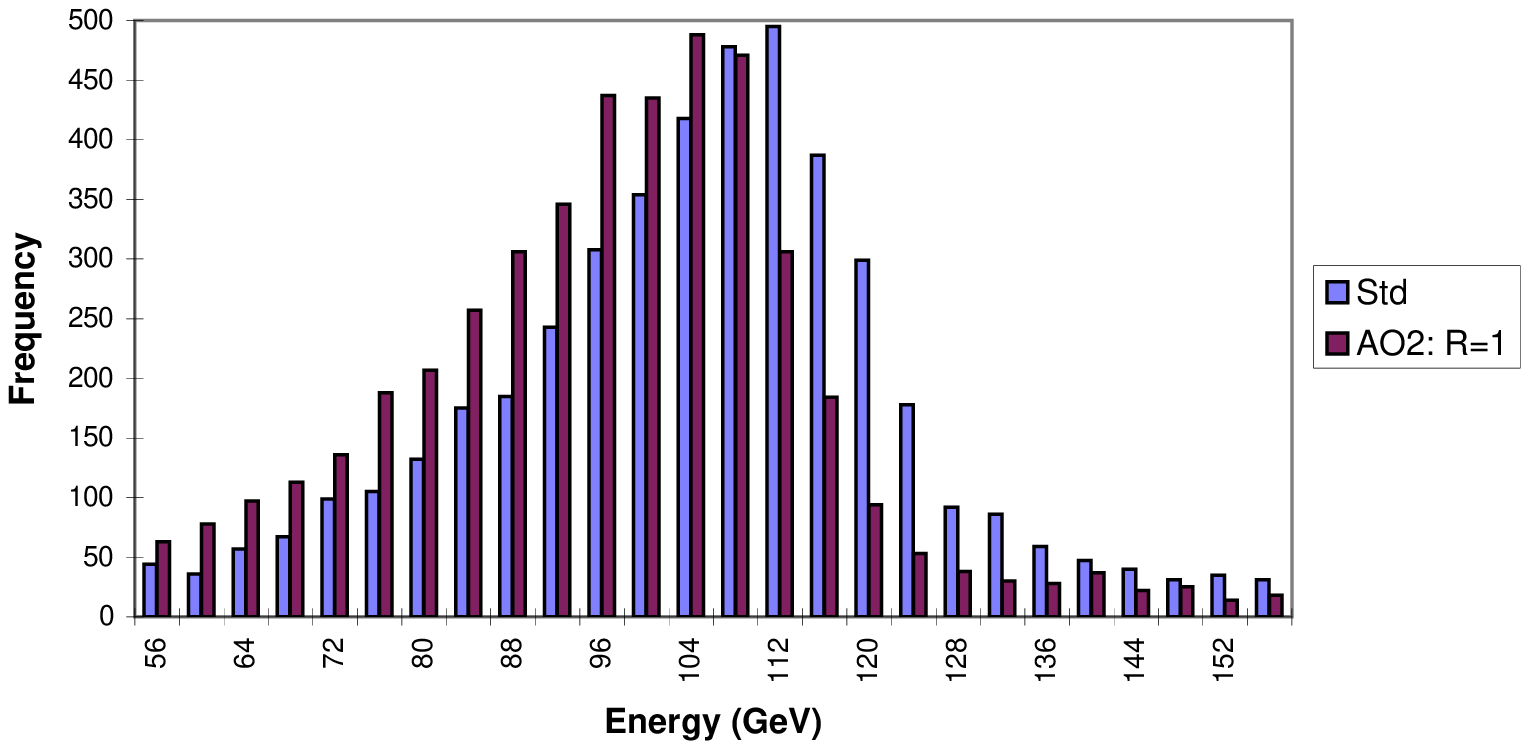}
\epsfbox[50 535 535 770]{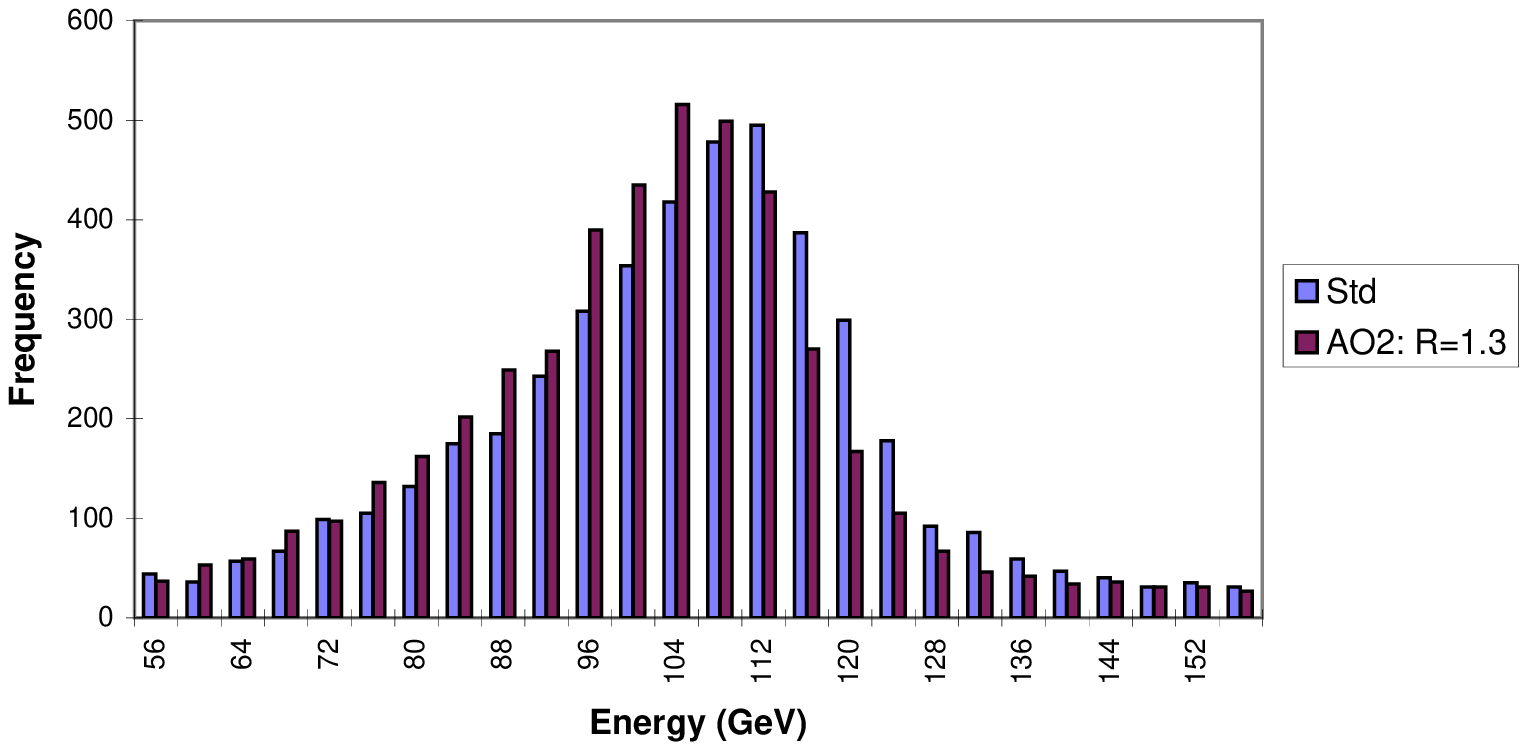}
\epsfbox[50 535 535 770]{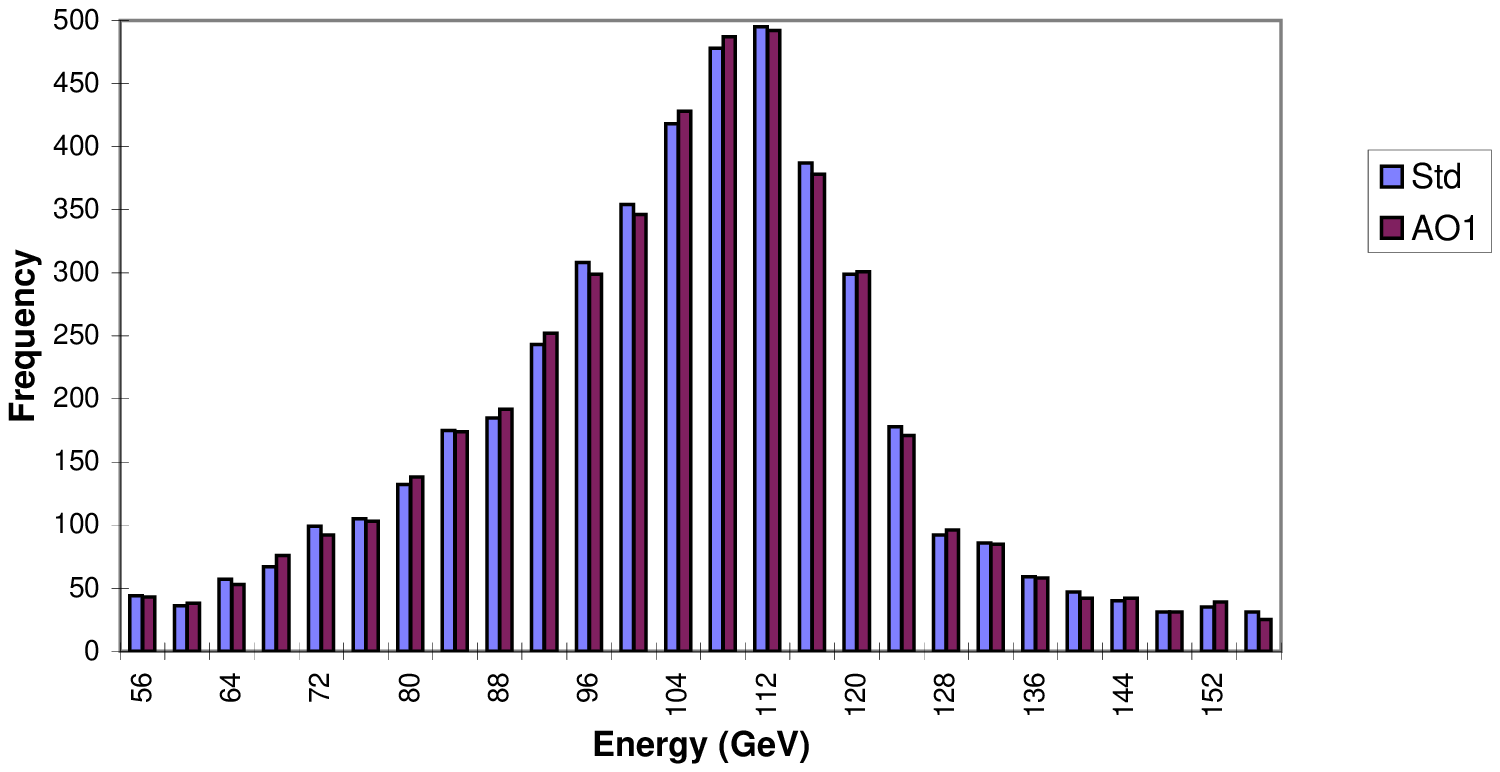}
\caption{Higgs masses were reconstructed through 5000 Higgs decays generated
by \HW.  The calorimeter threshold is 0.5 GeV.}
\label{fig_higgsm}
\end{figure}

Initially, we used a rather harsh threshold of 1 GeV to limit the
multiplicity of calorimeter cells in the final state.  This slightly
simplified situation is sufficient to show that the $AO_{1}$ algorithm and
standard $\kp$ algorithm reconstruct Higgs masses in a nearly identical
fashion, as illustrated in figure \ref{fig_higgsm}.

In order to reduce the energy lost through the high calorimeter
threshold, we also examined the case where the threshold was lowered to an
energy of 0.5 GeV, which practically eliminated the large mass offset.

As one would expect, there is a larger difference between the standard
$\kp$ algorithm and the $AO_{2}$ algorithm. The Higgs mass reconstructed by
the $AO_{2}$ algorithm with a cone size parameter $R=1$ is significantly
below the mass reconstructed by the standard algorithm.  Again, this is
due to the fact that combination is suppressed by the lower
effective cone size (see section \ref{sec_tube}).

Recall that at $R=1.3$, the amount of clustering
by the $AO_{2}$ algorithm was nearly equal to the amount of clustering with
the standard $\kp$ algorithm.  So it is of interest to see if the $AO_{2}$
algorithm reproduces results similar to the standard $\kp$ algorithm at
this value of $R$.  In fact, it does very nearly reproduce these results,
as shown in figure \ref{fig_higgsm}.  A summary of the reconstructed Higgs
masses is given in table \ref{tab_higgsm}. A simliar study using the cone algorithm yielded comparable results.  

\section{Conclusions}\label{sec_conc}
We have studied two possible modifications to the 
``standard'' \cite{CDSW,EllSop} $\kp$-clustering algorithm for hadron-hadron
collisions, both involving angular ordering, which has been found beneficial
for some purposes
in $\ee$ jet studies \cite{DLMW,MLS}. The first modified algorithm,
$AO_{1}$, simply changes the order of clustering by examining the
smallest-angle pair of remaining objects first.  The second, $AO_{2}$,
uses in addition a smoother measure of separation, analogous to that
in the {\tt LUCLUS} algorithm \cite{JETSET}, which has also been
recommended for $\ee$ physics \cite{MLS}.

We find that, in contrast to the $\ee$ case, these modifications
do not appear to provide any significant advantage over the
standard hadronic $\kp$-clustering algorithm.  No significant difference was
detected in the energy spectra, radial moment, or jet axes between the
standard algorithm and the $AO_{1}$ algorithm.  The
new separation measure in the $AO_{2}$ algorithm did provide
some differences.  However, these differences could mostly be
removed by simply increasing the cone size parameter, $R$.  The $AO_2$
algorithm may have less sensitivity to the calorimeter threshold and
the soft underlying event, but the differences are not very significant.

\acknowledgments 

A.T.P.\ acknowledges the financial support of the
Student Aid Foundation of Houston, TX, and Trinity College, Cambridge.




%
%



\begin{figure}
\epsfbox[50 535 535 770]{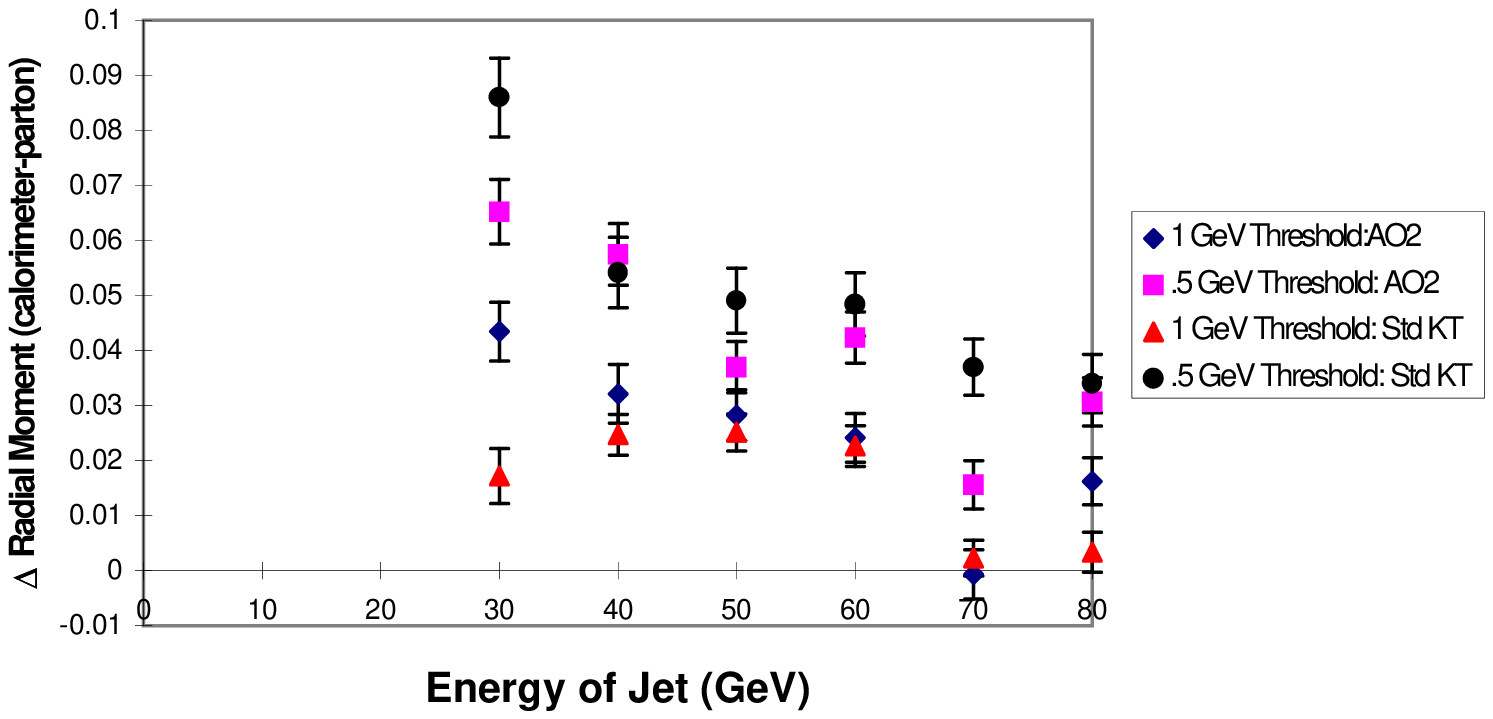}
\caption{A comparison between the standard $\kp$ and $AO_{2}$ algorithms at different
calorimeter thresholds.  Here we examine the difference in the radial moment
between the calorimeter and parton levels.}
\label{fig_thresh2}
\end{figure}

%
%

\begin{table}
\caption{Multiplicity of jets in a simple tube model. Errors are statistical.}
\label{tab_mult}
\begin{tabular}{||l|l||}  \hline \hline
Algorithm Type  &  Particles in Max $E_{T}$ jet \\ \hline
STANDARD $\kp$    & $1.726\pm.007$  \\  \hline
$AO_{1}$                  & $1.727\pm.007$  \\  \hline
$AO_{2}$                  & $1.446\pm.006$  \\    \hline \hline
\end{tabular} 
\end{table}

\begin{table}
\caption{Shift in Jet Axes. Errors are statistical.
Here calorimeter threshold=1GeV.}
\label{tab_shift}
\begin{tabular}{||l|l|l|l||}  \hline \hline
{\tt PTMIN}     &  Std. $\kp$ $\Delta R$ & $AO_{1}$ $\Delta R$ & $AO_{2}$
$\Delta R$\\ \hline
30 & $.109\pm.008$  &  $.110\pm.008$ & $.108\pm.009$\\ \hline
50 & $.070\pm.006$  &  $.071\pm.005$ & $.073\pm.006$\\ \hline
70 & $.062\pm.005$  &  $.062\pm.005$ & $.055\pm.005$\\ \hline 
90 & $.052\pm.004$  &  $.054\pm.005$ & $.050\pm.005$ \\ \hline \hline
\end{tabular} 
\end{table}

\begin{table}
\caption{Shift in Jet Axes. Errors are statistical. Here {\tt PTMIN} is
fixed at 50 GeV, and the sample size is 1000 events.}
\label{tab_calshift}
\begin{tabular}{||l|l|l|l||}  \hline \hline
threshold       &  Std. $\kp$ $\Delta R$ & $AO_{1}$ $\Delta R$ & $AO_{2}$
$\Delta R$\\ \hline
.5 GeV & $.070\pm.006$  &  $.071\pm.005$ & $.072\pm.006$\\ \hline
1 GeV  & $.072\pm.005$  &  $.070\pm.005$ & $.071\pm.006$\\ \hline \hline
\end{tabular} 
\end{table}

\begin{table}
\caption{An analysis of the sensitivity of the shift jet axis to the
underlying event. Errors are statistical. Here {\tt PTMIN} is fixed at
10 GeV, and the calorimeter threshold is 0.4 GeV}
\label{table_sue_shift}
\begin{tabular}{||l|l|l|l||}  \hline \hline
SUE status      &  Std. $\kp$ $\Delta R$ & $AO_{1}$ $\Delta R$ & $AO_{2}$
$\Delta R$\\ \hline
ON  & $.383\pm.019$  &  $.395\pm.020$ & $.401\pm.020$\\ \hline
OFF & $.344\pm.019$  &  $.336\pm.018$ & $.347\pm.019$\\ \hline \hline
\end{tabular}
\end{table}

\begin{table}
\caption{Reconstructed Higgs mass. Statistical errors only
(sample size 5000 events).}
\label{tab_higgsm}
\begin{tabular}{||l|l||}  \hline \hline
Threshold = 1 GeV & \\ \hline
Algorithm Type  &  Mean Higgs Mass  \\  \hline
$STANDARD$ $\kp$    & $101.1\pm.6$  \\  \hline
$AO_{1}$,           & $101.0\pm.6$  \\  \hline
$AO_{2}$, $R=1.3$   & $98.9\pm.6$   \\  \hline
$AO_{2}$, $R=1.0$   & $95.1\pm.6$   \\  \hline \hline
Threshold = 0.5 GeV & \\  \hline
Algorithm Type & Mean Higgs Mass \\ \hline
$STANDARD$ $\kp$  & $110.7\pm.7$  \\  \hline
$AO_{1}$,         & $110.6\pm.7$  \\  \hline
$AO_{2}$, $R=1.3$ & $107.0\pm.6$  \\  \hline
$AO_{2}$, $R=1.0$ & $101.1\pm.6$  \\   \hline \hline
\end{tabular} 
\end{table}


\begin{thebibliography}{99}
\bibitem{Snowmass}
J.E.\ Huth {\it et al.}, in {\em Research Directions for the Decade},
Proc.\ 1990 Summer Study on High Energy Physics, Snowmass, Colorado
(World Scientific, 1992), p.~134.
\bibitem{FlaMei}
For a summary of earlier efforts see:
B.\ Flaugher and K.\ Meier, in {\em Research Directions for the Decade},
Proc.\ 1990 Summer Study on High Energy Physics, Snowmass, Colorado
(World Scientific, 1992), p.~125.
\bibitem{SteWei}
G.\ Sterman and S.\ Weinberg, Phys. Rev. Lett.  {\bf 39}, 1436 (1977).
\bibitem{JETSET}
T.\ Sj\"ostrand, Comp.\ Phys.\ Commun.\ {\bf 28} 227 (1983).
\bibitem{JADE}
JADE Collaboration, W.\ Bartel et al., Z.\ Physik {\bf C 33} 23 (1986);
Phys. Lett. B {\bf 123} 460 (1993).
\bibitem{Durham}
Yu.L.\ Dokshitzer, contribution cited in report of Hard QCD Working Group,
{\it Proc.\ Workshop on Jet Studies at LEP and HERA, Durham, UK},
J.\ Phys.\ {\bf G 17} 1537 (1991).


\bibitem{CDOTW}
S.\ Catani, Yu.L.\ Dokshitser, M.\ Olsson, G.\ Turnock and B.R.\ Webber,
Phys.\ Lett.\ B {\bf 269} 432 (1991).

\bibitem{BKSS}
S.\ Bethke, Z.\ Kunszt, D.E.\ Soper and
W.J.\ Stirling, Nucl. Phys. {\bf B370}, 310 (1992), (erratum).

\bibitem{OPALcone}
OPAL Collaboration, R.\ Akers et al., Z.\ Physik {\bf C 63} 197 (1994).
\bibitem{ChaEll}
J.\ Chay and S.D.\ Ellis, Phys. Rev. D {\bf 55}, 2728 (1997).
\bibitem{EKS}
S.D.\ Ellis, Z.\ Kunszt and D.E.\ Soper, Phys. Rev. Lett. {\bf 62} 726 (1989); Phys. Rev. D. {\bf 40} 2188 (1989); Phys. Rev. Lett. {\bf 64} 2121, (1990).
\bibitem{GGK}
W.T.\ Giele, E.W.N.\ Glover and D.A.\ Kosower, Nucl. Phys {\bf B403} 633 (1993); Phys. Rev. Lett. {\bf 73} 2019 (1994).
\bibitem{radial}
W.T.\ Giele, E.W.N.\ Glover and D.A.\ Kosower, Phys. Rev. D. {\bf 57} 1878 (1998).
\bibitem{CDF}
CDF Collaboration, F.\ Abe {/it et al}., Phys. Rev. Lett. {\bf 70} 713 (1993), Phys. Rev. Lett. {\bf 77} 438 (1996).
\bibitem{D0}
D\/0 Collaboration, S.\ Abachi et al., Phys. Lett. B {\bf 357} 500 (1995);
V.D.\ Elvira, in {\it Proc.\ Rencontres de Physique de la Vallee d'Aoste,
LaThuile, Italy}, 1997.
\bibitem{CDSW}
S.\ Catani, Yu.L.\ Dokshitzer, M.H.\ Seymour and B.R.\ Webber,
Nucl. Phys. {\bf B406} 187 (1993).
\bibitem{EllSop}
S.D.\ Ellis and D.E.\ Soper, Phys. Rev. D. {\bf 48} 2160 (1993).
\bibitem{Seym93} 
M.H.\ Seymour, in {\it Proc.\ Les Arcs 1993, QCD and high energy hadronic
interactions}, p.141.
\bibitem{D094}
D0 Collaboration, K.C.\ Frame et al., 
FERMILAB-CONF-94-323G-E, presented at {\it 1994 Meeting
of the American Physical Society, Division of Particles
and Fields (DPF 94), Albuquerque, NM,} August 1994 (DPF Conf.\ 1994:1650).
\bibitem{D097}
D0 Collaboration, D.\ Lincoln et al., FERMILAB-CONF-94-323G-E,
to be published in {\it
Proc.\ 32nd Rencontres de Moriond: QCD and High-Energy Hadronic
Interactions, Les Arcs, France}, March 1997.
\bibitem{Seym94}
M.H.\ Seymour, Z. Physik {\bf C62} 127 (1994).
\bibitem{Pump}
J.\ Pumplin, Phys. Rev. D {\bf 55} 173 (1997). 
\bibitem{Seym98}
M.H.\ Seymour, Nucl. Phys. {\bf 513}, 269 (1998).
\bibitem{DLMW}
Yu.L.\ Dokshitser, G.D.\ Leder, S.\ Moretti and B.R. Webber,
JHEP {\bf 8} 001 (1997). 
\bibitem{MLS}
S.\ Moretti, L.\ L\"onnblad and T.\ Sj\"ostrand, preprint
RAL-TR-98-003A. 
\bibitem{BenMey}
S.\ Bentvelsen and I.\ Meyer, preprint CERN-EP/98-043. 
\bibitem{Feyntub}
R.P.\ Feynman, {\em Photon-Hadron Interactions} (Benjamin, 1972).
\bibitem{hadro}
B.R.\ Webber, in {\em Proc.\ Summer School on Hadronic Aspects of Collider
Physics, Zuoz, Switzerland,} August 1994. 
\bibitem{HW}
G.\ Marchesini, B.R.\ Webber, G.\ Abbiendi, I.G.\ Knowles, M.H.\ Seymour
and L.\ Stanco, Comput.\ Phys.\ Commun.\ {\bf 67} 465 (1992) 465.
\bibitem{CALO}
Calorimeter simulation {\tt CALSIM}, obtained from F.\ Paige.
\end{thebibliography}
\end{document}